# Attosecond angular streaking and tunnelling time in atomic hydrogen

U. Satya Sainadh[1], Han Xu[1], Xiaoshan Wang[2], Atia-Tul-Noor[1], William C. Wallace[1], Nicolas Douguet[3†], Alexander Bray[4], Igor Ivanov[5], Klaus Bartschat[3], Anatoli Kheifets[4], R. T. Sang[1] & I. V. Litvinyuk[1]

[1] *Australian Attosecond Science facility, Centre for Quantum Dynamics, Griffith University, Nathan, QLD 4111, Australia.*

[2] *School of Nuclear Science & Technology, Lanzhou University, Lanzhou, 730000, China.*

[3] *Department of Physics and Astronomy, Drake University, Des Moines, Iowa, 50311.*

[4] *Research School of Physics and Engineering, The Australian National University, Canberra, ACT 2601, Australia.*

[5] *Centre for Relativistic Laser Science, Institute for Basic Science, 123 Cheomdangwagiro, Buk-gu, Gwangju, 500-712, Korea*

† Present address: Department of Physics, University of Central Florida, Orlando, FL 32816.

**Tunnelling is one of the key features of quantum mechanics. A related debate, ongoing since the inception of quantum theory, is about the value, meaning and interpretation of 'tunnelling time'[1-5]. Simply put, the question is whether a tunnelling quantum particle spends a finite and measurable time under a potential barrier. Until recently the debate was purely theoretical, with the process considered to be instantaneous for all practical purposes. This changed with the development of ultrafast lasers and attosecond metrology[6], which gave physicists experimental access to the attosecond (1 as = $10^{-18}$ s) domain. It is at this time scale**

where most theoretically defined 'tunnelling times' belong. In particular, the 'attoclock'[7] technique was used to probe the attosecond dynamics of electrons tunnelling out of atoms interacting with intense laser fields. Although the initial attoclock measurement[7-10] hinted at instantaneous tunnelling, later experiments[11,12] contradicted those findings, claiming to have measured finite tunnelling times. In each case these measurements were performed with multi-electron atoms. For such targets accurate theoretical modelling is not available, thereby complicating the interpretation of the ionization dynamics. Atomic hydrogen (H), the simplest atomic system with a single electron, can be 'exactly' (subject only to numerical limitations) modelled using numerical solutions of the three-dimensional time-dependent Schrödinger equation (3D-TDSE) with measured experimental parameters. Hence it acts as a convenient benchmark for both accurate experimental measurements and calculations[13-15]. Here we report the first attoclock experiment performed on H using a 'Reaction Microscope'[16] (REMI) and 770 nm, 6 fs pulses (FWHM) with peak intensities of 1.65 - 3.9 $\times 10^{14}$ W/cm$^2$ . We find that our experimentally determined offset angles of the photoelectrons are in excellent agreement with accurate 3D-TDSE simulations performed using the Coulomb potential with our experimental pulse parameters. The same simulations with a short-range Yukawa potential result in zero offset angles for all intensities. We conclude that the offset angle measured in the attoclock experiments originates entirely from electron scattering by the long-range Coulomb potential with no contribution from tunnelling time delay. Thus we confirm that, in atomic H, tunnelling is instantaneous within our experimental and numerical uncertainty. This puts an upper limit of 1.8 attoseconds on possible delays due to tunnelling. The result is in agreement with the recent theoretical findings[17] and in effect rules out all commonly used 'tunnelling times'[18] from being interpreted as time spent by an electron under the potential barrier[19].

In a photoionization process, an electron is liberated from a bound state into a continuum state when one or more quanta of electromagnetic radiation (photons) are annihilated (absorbed). The process is inherently quantum mechanical, and its understanding and mathematical description go to the very core of the wave-particle duality. Einstein's attempt to understand and explain photoionization in the photoelectric effect laid the foundation for one of the most successful theories of our understanding of nature, namely quantum mechanics. The H atom has played a major role in that success because of its simplicity. It serves as a physical two-body system that can be treated analytically (in the non-relativistic approximation) and yields solutions in closed form. It has been long known that atomic physics experiments performed with H could be used as a point of reference for our understanding of the inherently complex dynamics of light-matter interactions, in particular strong-field physics[13-15]. As simulations of the strong-field processes remain computationally demanding, it is still very important to validate the accuracy of those numerical models against precise measurements. The present work was motivated by an earlier theoretical study of the attoclock in H[20], which made a compelling, albeit purely theoretical, argument in support of instantaneous tunnelling. Here we validate that argument by experimentally confirming the numerical predictions of the 3D-TDSE simulations in the context of the attoclock. Our work also removes the possibility of electron-electron interactions influencing the interpretation of attoclock measurements for tunnelling time determination.

The experimental technique of attosecond angular streaking, or attoclock, utilizes a nearly-circularly-polarized few-cycle infra-red (IR) pulse such that the evolving $\boldsymbol{E}(t)$ field vector, rotating by 360°, maps time to angle in the polarization plane (see figure 1). That single pulse provides both the ionizing radiation and the streaking field, therefore making the technique self-referencing. It was envisaged based on a two-step process of the simpleman's model[21,22], where the first step is the quantum process of electron tunnelling through the suppressed Coulomb barrier in the presence of a strong external field, $E(t)$ and the second step assumes a classical description of the electron-in-an-electromagnetic-field (streaking) from the instant it appears in the continuum whilst neglecting the Coulomb potential of the parent ion. The very significant non-linearity of tunnelling ionization ensures that the ionization rate peaks when the $E(t)$ field reaches its maximum. The streaking field drives the ionized electron in such a way that its final momentum (following the interaction with the pulse) is equal to the negative instantaneous value of the vector potential, $A(t)$ of the streaking field at the moment of ionization. Hence, the technique involves a well-defined 'time-zero', which in this case is the direction of the maximum field, and also naturally encodes the information on the instant of ionization, the 'tunnel exit'; thus providing the information on any possible tunnelling delays.

Although the angular streaking works best with circularly polarized few-cycle pulses, there is an experimental issue with determining the angle at which the electric field (and related to it tunnelling ionization probability) reaches its maximum. That angle depends on the carrier-envelope phase (CEP) of such pulses. Presently the best stabilization techniques achieve CEP noise of about 100-150 mrad, corresponding to about 7° or 50 attosecons uncertainty in angle/time measurement, which is already comparable to, or exceeds the expected tunnelling times. However, this experimental issue is resolved by using slightly elliptically polarized light pulses. Even for an ellipticity of 0.88 (very close to circular) without CEP stabilization, the electric field will reach its maximum when it points along the major axis of the polarization ellipse[8], and the direction of the electric field can be determined with high precision using basic polarimetry.

In our experiment we used laser pulses centred at 770 nm with 6 fs duration and ellipticity of 0.84±0.01. The atomic H jet, generated using a discharge source with a dissociation ratio of 50%, is integrated with a REMI. The reader is referred to Methods and Extended Data (Fig. 1E) for further details of the experimental set-up. Once the polarization ellipse is defined, we ionize H from the atomic gas jet and each electron-proton pair is detected in coincidence. In order to confirm that the electron-proton pair originates from the same H atom, we ensure that the ionization events per pulse are low enough to suppress the false coincidence probability to be less than 4%. The peak intensity of the laser pulses was varied from $1.65×10^{14}$ W/cm$^2$ to $3.9×10^{14}$ W/cm$^2$. Figure 2 illustrates how the angular offsets were extracted from our experimental data for the intensity $1.95×10^{14}$ W/cm$^2$.

The attoclock observables from the experiment are directly compared to the *ab-initio* simulations provided by two independent theoretical groups. The full solutions of the 3D-TDSE, calculated accurately within the non-relativistic framework using the electric dipole approximation, generate photoelectron momentum distributions (PMD) projected on the polarization plane with the same pulse parameters as were used in the experiments. Since the CEP was not stabilized in our experiments, the simulations average the momentum distributions over eight CEP values ranging from 0 to $2\pi$ in steps of $\pi/4$. These calculated momentum spectra were then analysed using the same methods as in the case of their experimental counterparts. The angular offsets obtained from theory are compared to the experimental results (see figure 3). The error bars associated with the theoretical data are due to uncertainties in our fitting procedure used to determine the angular offsets.

The experimental results are in excellent agreement with those of the theoretical simulations from the two independently developed TDSE codes, which are also in mutual agreement. We see a trend of the angular offset decreasing as the field strength increases. The Coulomb potential by itself is known to produce angular offsets in momentum distributions of electrons ionized by an elliptically polarized light even in the absence of any tunnelling delay. Therefore it is critical to disentangle the effect of Coulomb potential from that of the tunnelling delay. While the Coulomb potential cannot be removed or replaced by a different potential in experiment, this can easily be done in theoretical simulations.

We disentangle the effect of the Coulomb potential by extracting angular offsets from PMDs simulated using a short-range Yukawa potential with the same experimental pulse parameters. The Yukawa potential is modelled as $U_Y = -Z/r \, e^{-r/a}$ with parameters $Z = 1.908$ and $a = 1$, such that the ground-state energy remains the same (see Extended Data Fig. 3). The computations with the Yukawa potential show a zero angular offset (within our numerical and fitting uncertainty) for all intensities. That leads to the definitive conclusion that for atomic H attoclock the offset angles originate entirely from Coulomb scattering of the ejected electron with no contribution from any tunnelling delay. Consequently, our results are consistent with zero time delay between the peak of the electric field and the appearance time of the continuum electron measured by the attoclock, corresponding to instantaneous tunnelling.

Based on our estimated experimental and numerical uncertainties we can put an upper bound on the tunnelling time delays by considering the angular offset, in the case of the Yukawa potential, that has deviated from the zero reference line the most. We estimate that the angular offset cannot exceed 0.25° for the Yukawa potential, which corresponds to the maximum possible tunnelling time delay of 1.8 attoseconds. That is substantially less than values of any commonly used theoretical definitions of tunnelling times[18] (i.e., Keldysh time, Büttiker-Landauer time, Eisenbud-Wigner time, Pollack-Miller time, Larmor time, Bohmian time) which are all in the ten to few hundreds attoseconds range. Thus we effectively rule out all of these 'tunnelling times' from being interpreted as the time spent by a quantum particle under a potential barrier. It is also likely that any experimental search for a finite tunnelling time will have to explore the zeptosecond ($10^{-21}$ s) time domain.

We believe that through both precise measurements and high-accuracy *ab initio* simulations, properly validated by mutual comparison, the issue of any possible tunnelling delays was addressed meaningfully and unambiguously by our study. We anticipate our results to have strong implications, as we have completely excluded the issues of any multi-electron effects that were present in other experiments. High-precision experiments with a benchmark system such as H open the way towards accurate measurements of photoionization delays for various multi-electron atoms and molecules. Those measurements will provide important information on ultrafast electron dynamics, in particular on electron-electron correlation and interactions. Finally, the tunnelling events in strong field ionization of H are only as 'instantaneous' as the electron wave function collapse, with which appearance of continuum electrons is associated in the orthodox interpretation of quantum mechanics. Therefore, future measurements of tunnelling delay times with greater accuracy in zeptosecond or sub-zeptosecond domain may open an intriguing possibility of observing the dynamics of the wave function collapse itself, thus breaching the limits of the Copenhagen interpretation.

**Methods**

**Experimental details and set-up.** We create an atomic H jet using a radio-frequency (RF) discharge tube based on the design[23] that dissociates hydrogen molecule via electron impact. Hydrogen gas from the cylinder is sent through a Pyrex glass tube mounted inside a quarter-wave helical resonator, which is powered with an amplified RF-signal at 75 MHz. The standing wave formed in the tube strikes a discharge, generating a plasma which dissociates molecular hydrogen. The discharge produces a beam of atomic hydrogen with a constant dissociation fraction defined by the number density of atomic and molecular hydrogen as

$$\mu = \frac{[\text{H}]}{[\text{H}] + 2[\text{H}_2]} .$$  (1)

In the experiment, we generated H with a dissociation fraction of 50%. Further details on its construction, operation, characterization and optimization are available at[24]. Before it reaches the interaction region, the hydrogen beam passes between two deflector plates used to remove any charged particles from the beam by applying a constant electric field of 26 V/cm over 6 cm length. That electric field also serves to quench any metastable hydrogen atoms (mostly H 2s) produced during dissociation (see Supplemental Section for details).

We generate 6 fs pulses (FWHM) around 770 nm central wavelength from the commercially available 'FEMTOPOWER Compact Pro CE-Phase' laser system that has a repetition rate of 1 kHz. Ions and electrons formed by the ionization of hydrogen atoms interacting with these laser pulses are detected in coincidence in the Reaction Microscope (REMI) apparatus. We ensure that less than one ionization event occurs for every laser pulse and analyse only the electrons coincident with very low-energy protons characteristic of atomic H ionization.

**Analysis of data and results.** The projected PMD in the polarization plane ($P_x$-$P_z$) is divided into optimal polar angular bins $\Delta\theta$, where $\theta$ defined as $\tan^{-1}(P_z/P_x)$ is the streaking angle. The distribution acquired by radially integrating the counts in each bin, $f(\theta)$ is plotted against the streaking angle and is then fitted using a double Gaussian function of the form:

$$f(\theta) = \sum_{i=1}^{2} a_i e^{-\left[\frac{\theta - b_i}{c_i}\right]^2} \qquad (2)$$

using a least-squares fitting routine. The values of $b_i$ gives the direction of most probable ejection ($\theta_{streak}$); and the fitting procedure provides a 95% (2σ) level confidence bounds of the fit from which the standard error, $\delta\theta_{streak}$ is inferred.

The angular offset is defined as the angular difference in the direction of most probable photoelectrons emitted and that of the maximum field. However the maximal field is determined using polarimetry in the polarizer's reference frame and the PMD is measured in REMI's frame of reference. In order to have an absolute comparison of the measured angles in the polarization measurement and the angular distributions, we use a polarizer to convert the laser pulse into linear polarization before sending it into REMI to ionize atomic H. Since with linearly polarized field, the electrons are mainly emitted along the laser polarization axis, which is parallel to the polarizer's optical axis, we calibrate the angle of the polarizer's optical axis with the angle of the peak PMD measured in the REMI's coordinate system (see Extended Data Fig. 2E). The angle of peak PMD is extracted using the above described double Gaussian fit routine and a relative offset of $\theta_{sys}$ with an error of $\delta\theta_{sys}$ between the two coordinate systems is determined. This enabled us to work in the same frame of reference such that the relation is $\theta = \theta_{pol} - \theta_{sys}$, where $\theta_{pol}$ is the angle of the polarizer's optics axis.

We performed optical polarimetry measurements to determine the ellipticity and then used the H ion yields with the calibrated angle to find the major axis of the polarization ellipse. Given the highly non-linear dependence of the ionization probability as a function of the electric field, this method provides a more accurate determination of the polarization ellipse orientation for day-to-day measurements. We extracted the ellipticity of the polarization ellipse through fitting the laser power $P$ measured for polarizer angle $\theta$ with the function

$$P(\theta) = A^2 \sin^2(\theta - c) + B^2 \cos^2(\theta - c). \qquad (3)$$

Here $B/A$ gives the ellipticity, manifested as the modulation depth, while $c$ gives the angle at which the polarization ellipse, i.e. its major axis, was oriented. The ellipticity is defined as $\varepsilon$ = major axis/minor axis (of the **E** field, which is proportional to Intensity$^{0.5}$), such that it spans the range from 0 to 1 corresponding to linear and circular polarization, respectively. The error in ellipticity is calculated as $\delta\varepsilon/\varepsilon = \{(\delta B/B)^2 + (\delta A/A)^2\}^{1/2}$, where $\delta B$ $(\delta A)$ are the standard errors extracted from the fits. To find the major axis we used the fitting function $e^{-P(\theta)}$ that fits the atomic H ion yields.

The error in the determining the relative offset in the calibration procedure contributes to the error in the fit, $\delta\theta_{fit}$ to give us $\delta\theta_{ellipse} = \sqrt{\delta\theta_{sys}^2 + \delta\theta_{fit}^2}$. Having both the measurements precisely taken in the same frame of reference enables the angular offset and its corresponding error to be determined via

$$\theta_{offset} = \theta_{streak} - \theta_{ellipse} - 90^0 \qquad (4)$$

$$\delta\theta_{offset} = \sqrt{\delta\theta_{ellipse}^2 + \delta\theta_{streak}^2} \qquad (5)$$

where and $\theta_{streak}$ and $\theta_{ellipse}$ are defined as the direction of the most probable photoelectron ejection and orientation of the major axis in the polarization plane, respectively. Here the streaking angle of the light pulse is -90°, since the momentum of the streaked electrons is determined by the vector potential of the light field of the pulse, which itself lags the electric field by 90°.

**Numerical Simulations.** The numerical methods employed in two sets of calculations to solve the 3D TDSE are conceptually similar, both relying on spherical-harmonics expansions of the wave-function to represent its dependence on the angular variables and treating the radial variable by discretizing the TDSE on a grid. Both groups used the Matrix Iterative Method[25] to propagate the initial state in time using the velocity gauge in the electric dipole approximation. Detailed descriptions of the numerical techniques can be found in[26,27]. Careful checks were performed to ensure that convergence with respect to the parameters defining the accuracy of the calculation (e.g., the number of partial waves in the expansion as well as the step sizes on the space-time grid) was achieved. For the peak intensity of $1.65 \times 10^{14}$ W/cm$^2$, the results from the two groups were compared to ensure that the independent implementations of the computational techniques gave same results within the error bar due to the uncertainty of the fitting procedure used to determine the angular offset. The maximum orbital angular momentum $l_{max}$ needed in the partial-wave expansion was 40 for $1.4 \times 10^{14}$ W/cm$^2$ and 100 for $3.9 \times 10^{14}$ W/cm$^2$, respectively. Specifically, we set the components of the vector potential $\boldsymbol{A}$(t) for a pulse with ellipticity $\varepsilon$ as

$$A_x(t) = -\frac{E_0}{\omega}\, f(t)\, \frac{\varepsilon}{\sqrt{1+\varepsilon^2}} \cos(\omega t + \varphi)$$

$$A_y(t) = -\frac{E_0}{\omega}\, f(t)\, \frac{1}{\sqrt{1+\varepsilon^2}} \sin(\omega t + \varphi)$$

$$A_z(t) = 0$$

The envelope function $f(t)$ was a Gaussian ramped on and off over three optical cycles each, respectively. The electric field was obtained as $\boldsymbol{E}(t) = -d\boldsymbol{A}(t)/dt$. Finally, the carrier envelope phase $\varphi$ was varied in steps of $\pi/4$ and the results were averaged.

'**Supplementary Information** accompanies the paper on **www.nature.com/nature**.'

**Acknowledgements** The experiments were performed at the Australian attosecond science facility at Griffith University, which was partially supported by the ARC. U.S.S., A.T.N, and X.W. were supported by GUIPRS. H.X. was supported by an ARC Discovery Early Career Researcher Award (DE130101628). The work of N.D. and K.B. was supported by the United States National Science Foundation under grant No. PHY-1430245 and the XSEDE allocation PHY-090031. Their calculations were performed on SuperMIC at the Center for Computation & Technology at Louisiana State University.

**Author Contributions:** U.S.S, X.W and W.C.W. integrated the hydrogen source on to REMI as part of the experimental set-up. U.S.S with assistance of H.X. and A.T.N performed the experiments. U.S.S. and H.X. analysed the experimental data and post-processed theoretical data. Theoretical predictions were provided by A.B., A.K., I.I., N.D. and K.B. The project was supervised by R.T.S and I.V.L. All authors discussed the results and contributed to the manuscript.

**Author Information:** Reprints and permissions information is available at www.nature.com/reprints. The authors declare no competing financial interests. Readers are welcome to comment on the online version of the paper. Correspondence and requests for materials should be addressed to Han Xu (h.xu@griffith.edu.au) or R.T. Sang (r.sang@griffith.edu.au.) or I.V. Litvinyuk (i.litvinyuk@griffith.edu.au).

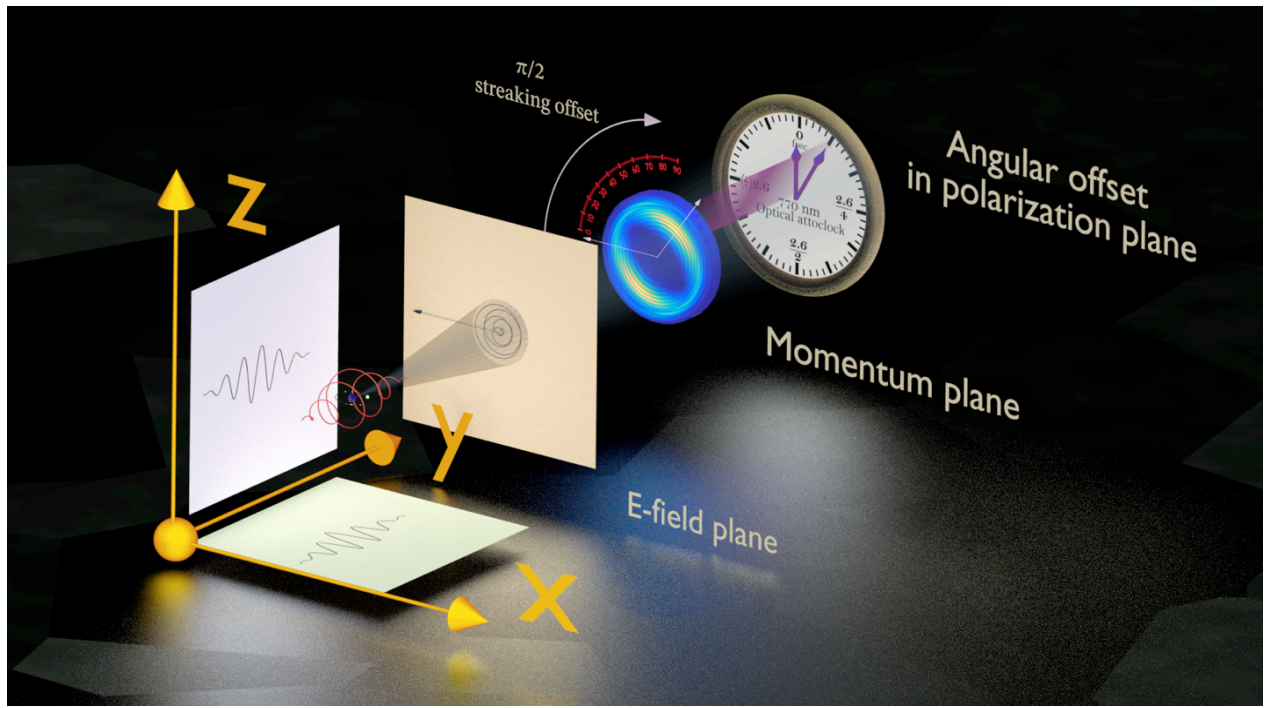

Figure 1: **Attosecond Angular Streaking**: The temporal evolution of the field vector of a typical 770 nm pulse of 6 fs duration provides a unique direction of the maximal field vector in the polarization plane. A strong electric field can bend the atom's binding potential, allowing for the electron to tunnel out. For a circularly polarized field the vector-potential trails the rotating electric field vector by 90°. Consequently, upon ionization, the electron will be emitted perpendicularly to the instantaneous direction of the electric field at the moment of ionization. However, any delay between the electron's exit and an independently measured maximum of the **E** field, where the tunnelling probability is maximal, manifests itself as an angular offset in the photoelectron momentum. As the Coulomb potential introduces an angular shift to the electron momentum in a direction dictated by the right- or left-hand circularly-polarized streaking field, any additional delay due to the time spent by the electron under the potential barrier will result in a greater angular shift in the electron's momentum and thus could be measured experimentally. As a full 360° revolution of the electric field vector for a 770 nm pulse is completed in 2.6 fs, a 1° in offset angle is equivalent to a 2.6 fs/360° ≈ 7.13 attoseconds in the electron ionization delay. Hence, by measuring the 3D PMD one can reconstruct the distribution of ionization time delays with respect to the time when the electric field reaches its maximum.

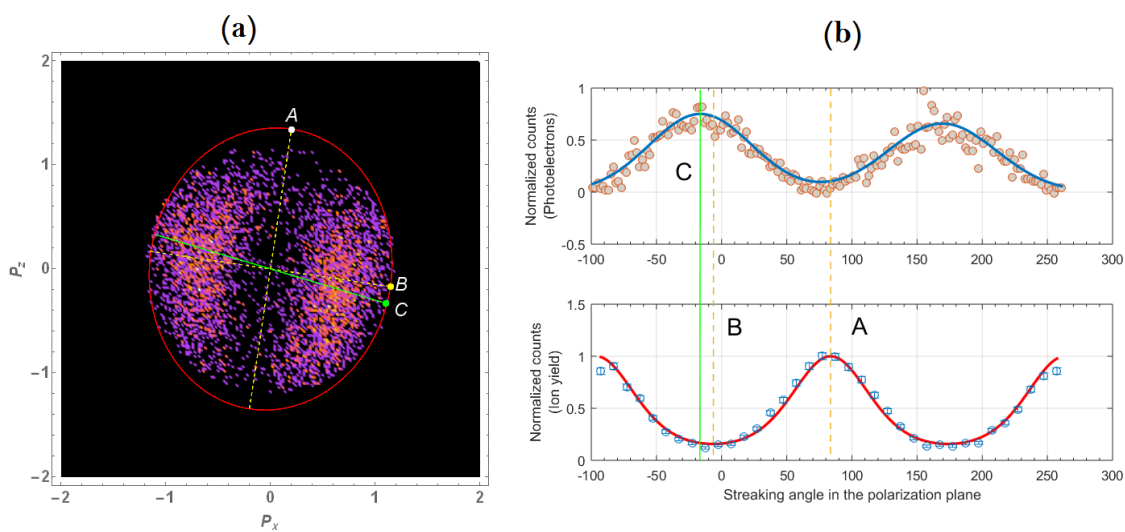

Figure 2: **Angular Offsets**. The presented data are for a peak intensity of 1.95 ×10$^{14}$ W/cm$^2$. The major axis of the polarization ellipse defines the direction of the peak electric field and can be determined by basic polarimetry after careful calibration (see Methods). **(a)** Experimental data of momentum distribution of photoelectrons in the polarization plane. **A** corresponds to the peak electric field and **B** & **C** are the expected and measured peaks of PMD in the polarization plane. **(b)** (Top) The cumulative photoelectron signal in the polar angular bins of 2° each in the polarization plane. A double Gaussian function is fitted to determine position of each peak. (Bottom) Atomic H-ion yield data as a function of calibrated polarizer angle is fitted using the function e$^{-P(\theta)}$ (see Methods) to determine the major axis of the polarization ellipse. The errors in the fit are determined by the confidence bounds of the various fit parameters involved in the fit function. The points **A**, **B**, **C** in **(a)** are also shown in **(b)** to illustrate the measured angular offsets. The angular difference of 10.87°(±1.42°) between **B** and **C** is the measured angular offset.

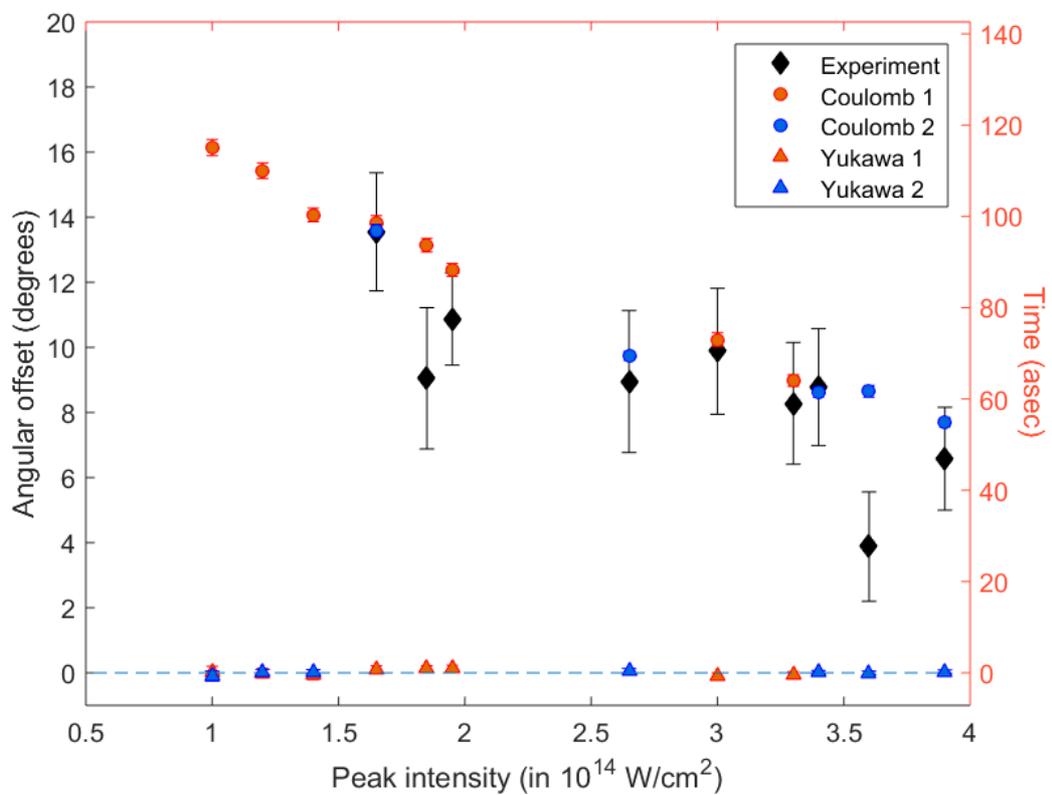

Figure 3: **Results**. The experimental observations are compared to *ab-initio* 3D-TDSE simulations with both Coulomb potentials, provided by two independent groups marked as 1 and 2. To disentangle the effects of the Coulomb potential on the continuum electron, we also include the TDSE simulations for a Yukawa potential. The same extraction procedure was used to determine the offset angles from experimental results and theoretical simulations for both Coulomb and Yukawa potentials. Our numerical experiment demonstrates that the observed angular offsets are entirely due to the electron scattering by the long-range Coulomb potential of the ion.

**Extended data**

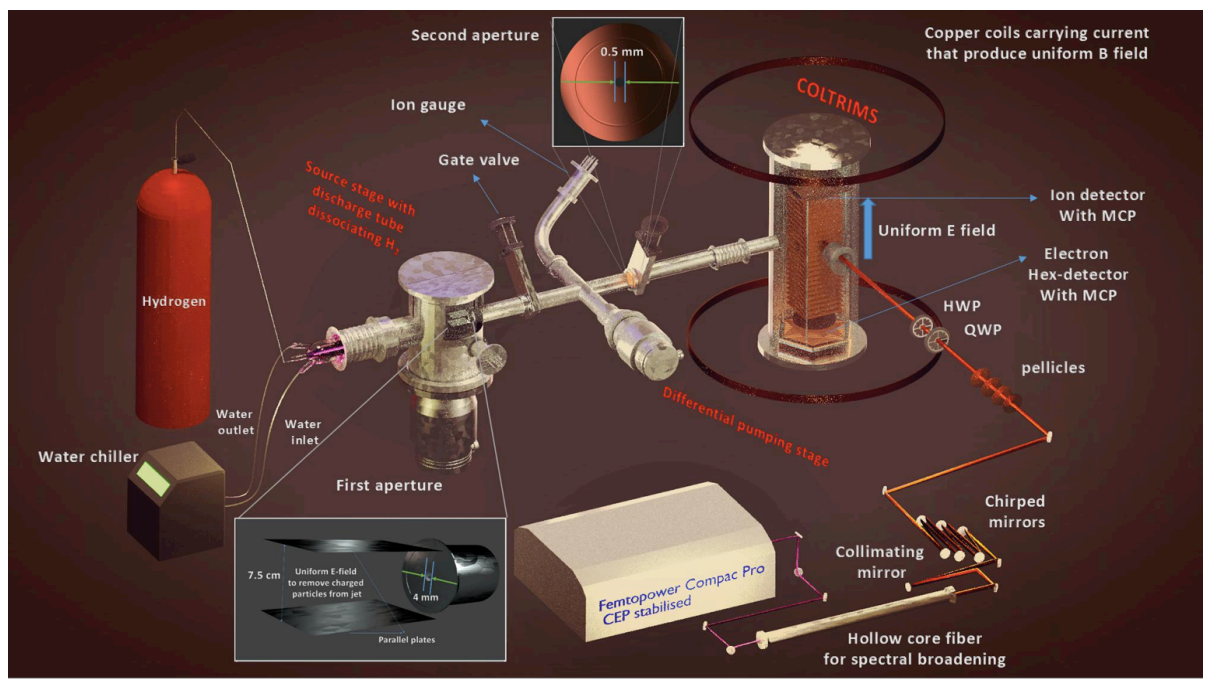

Figure 1E: **Experimental set-up**. As part of the atomic beamline the water-cooled Pyrex discharge tube dissociating $H_2$ to H is driven with an optimized load of hydrogen gas ($H_2$), placed under a vacuum of $10^{-5}$ mbar in a source chamber. Any charged species coming from the jet are expelled using a uniform electric field before passing through an aperture to the next stage. A differential pumping stage, typically maintained at $10^{-7}$ mbar pressure, is employed to ensure no possible recombination of H atoms before they finally enter the REMI, commonly also known as 'Cold target recoil ion momentum spectroscope' (COLTRIMS), through a 0.5 mm aperture as a supersonic jet. The few-cycle laser pulses then pass through a series of pellicle beamsplitters (used for varying the intensity) and ultra-broadband waveplates becoming elliptically polarized pulses that interact with H in the REMI. Fragments after the photoionization events accelerate in the uniform electric and magnetic fields of the REMI and finally get detected on the position sensitive detectors. The added information of position with time-of-flight enables us to reconstruct the 3D-momentum distribution of the fragments, soon after the interaction with the laser.

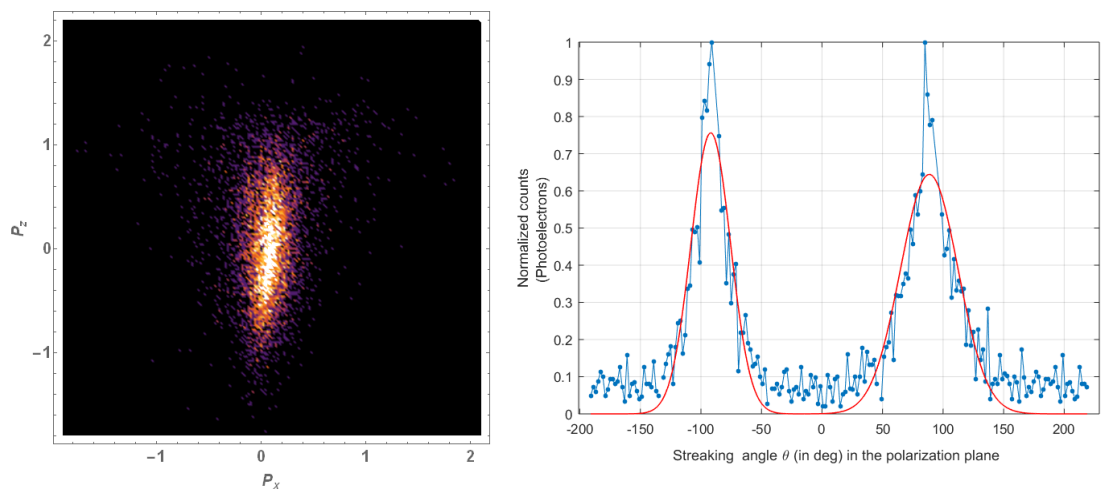

Figure 2E: **Calibration of reference frame**. The elliptically polarized few-cycle pulse is passed through a polarizer, with its optical axis aligned to the reading 90° on it, which makes it linear, and then made to interact with the atomic jet inside the REMI. The PMD in the polarization plane is then plotted as a function of streaking angle. The orientation of the field is found by determining the peaks of the distribution, thereby giving us a systematic offset between the frames of the polarizer and the REMI. The polarization is chosen parallel (in this case vertical) to the time-of- flight axis due to better momentum resolution. A systematic offset of ~3° was measured and used to calibrate our measured offsets and polarimetry.

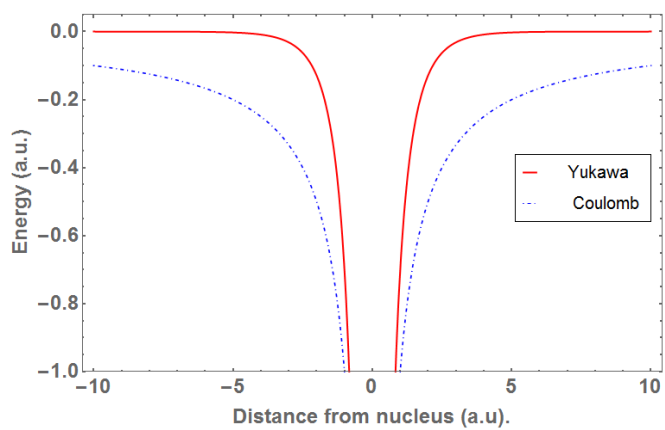

Figure 3E: **Yukawa Potential**. A short-range Yukawa potential was used to investigate the effects of the electric field of the resultant ion-core on the electron. The Yukawa potential is of the functional form $U_Y = -Z/r \, e^{-r/a}$ with parameters $Z$ = 1.90831 and $a$ = 1. It goes to zero quickly in comparison to its Coulomb counterpart. The parameters were chosen to retain the ground-state energy of H.

**Supplementary Information**

**Excited states of H**

Discharge dissociation of $H_2$ by electron impact can produce electronically excited hydrogen atoms via dissociation of excited hydrogen molecules. Most of those excited states will decay radiatively into 1s or 2s states within few nanoseconds. The 2s state is metastable (its radiative one-photon transition to the ground state is forbidden by the selection rules) with field-free lifetime of 125 milliseconds determined by the rate of the two-photon transition. However, that restriction is lifted by an external electric field which couples 2s and 2 p states, thus effectively quenching the 2s.

The lifetime of H 2s as a function of external electric field $E$ was calculated by Bethe and Salpeter (Bethe, H.A. & Salpeter, E.E., "Quantum Mechanics of One- and Two-Electron Atoms", Academic Press Inc., New York, 1957, Sec. 67) and the theoretical formula was experimentally verified by Sellin (Sellin, I.A. "Experiments on production and extinction of 2s states of the hydrogen atom", *Physical Review* **136**, A 1245 (1964)). That lifetime is given by

$$\tau(E) = \tau(2p)\{1 + \frac{\delta^2}{[1-(1+\delta^2)^{1/2}]^2}\}, \qquad \delta = \frac{2\sqrt{3}Eea_0}{L}$$

where $\tau(2p)$ is the lifetime of the 2p state, $L$ is the Lamb shift and $a_0$ is the Bohr radius.

For the electric field of 23 V/cm present inside of our spectrometer the lifetime of H 2s is less than 700 ns. It is even less for the field of 26 V/cm which exists between the 6 cm deflector plates used to remove charged particles from our beam. For a typical beam velocity of 2500 m/s hydrogen atoms will travel less than 2 mm during H 2s lifetime, with H 2s population reduced by a factor of $e$ along the way. As the atoms need to cover 10 cm through the electric field (6 cm between the deflector plates and 4 cm inside the spectrometer) to reach the laser focus, the population of H 2s will be reduced by the factor of at least $e^{50}$ or $10^{22}$ to essentially zero.

It is easy to confirm the absence of H 2s in the interaction region, due to their electron momentum distributions being very distinct from those of H 1s. We calculated those momentum distributions for the same pulse parameters using the same numerical simulation methods (see figure 1S). Due to its much lower ionization potential H 2s is ionized early in the pulse, so that the electrons gain less energy from the field. In addition to lower energy, angular distribution of the H 2s electron is nearly uniform. Even if significant fraction of the signal came from H 2s, the measured offset angles would not have been affected – only the modulation depth of the angular distribution would be diminished. However, the modulation depth measured in our experiments agrees with theoretical prediction for pure H 1s, indicating that there is very little, if any, H 2s in our target gas. Furthermore, we calculated the offset angles with the energy filter removing H 2s electrons imposed and obtained the same values within the error bars as before.

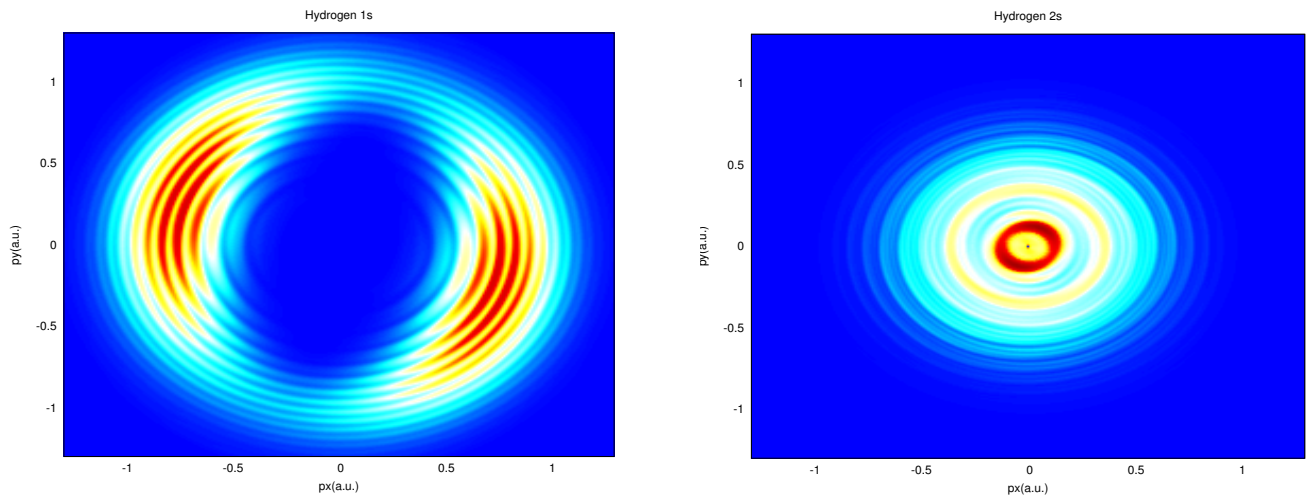

Figure 1S: **Simulated momentum distributions for 1s and 2s states of H.** Simulations performed for peak intensity of $1.9 \times 10^{14}$ W/cm$^2$ are shown for initial states 1s (left) and 2s (right). It is clear that the metastable H 2s is characterised by less energetic electrons with nearly uniform angular distribution.